\def\chisq{$\chi^{2}_{\nu}$}\def\minone{$^{-1}$}\def\mintwo{$^{-2}$}
\def\sqiggt{\hbox{\rlap{\lower.55ex \hbox {$\sim$}}
\kern-.3em \raise.4ex \hbox{$>$}\,}}
\def\sqiglt{\hbox{\rlap{\lower.55ex \hbox {$\sim$}}
\kern-.3em \raise.4ex \hbox{$<$}\,}} 
\def\kev{\,ke\kern-.1em V} \def\ev{\,e\kern-.1em V} \def\kalpha{K$\alpha$}
\def\sqig{$\sim\,$} \def\etal{et\,al.} \def\msun{M$_{\scriptstyle\odot}$} 
\def\up#1{$^{\mbox{{\scriptsize #1}}}$}  
 \def\pten#1{$\times10^{#1}$}
 \def\kmps{km\,s$^{-1}$} 
\def\asca{{\sl ASCA}}\def\rxs{RX\,1712--24}
\documentstyle{mn}
\title[Iron line widths]
{Iron \kalpha\ line widths in magnetic cataclysmic variables}
\author[C.~Hellier \etal]{Coel Hellier,$^{1}$ Koji Mukai$^{2}$ and
J.~P.~Osborne$^{3}$\\
$^{1}$Department of Physics, Keele University, Keele, Staffordshire, ST5 5BG\\
$^{2}$Laboratory for High Energy Astrophysics, Code 660.1, NASA/GSFC,
Greenbelt, MD 20771, USA\\
$^{3}$Department of Physics and Astronomy, University of Leicester,
University Road, Leicester LE1 7RH}

\date{ }
\begin{document}
\maketitle
\begin{abstract}
Following a recent report that AO~Psc has broad iron \kalpha\ emission lines
we have looked at the \asca\ spectra of 15 magnetic cataclysmic variables.
We find that half of the systems have \kalpha\ lines broadened by \sqig200
\ev, while the remainder have narrow lines. We argue that the Doppler
effect is insufficient to explain the finding and propose that the lines
originate in accretion columns on the verge of optical thickness,
where Compton scattering of resonantly-trapped line photons broadens
the profile. We suggest that the broadening is a valuable diagnostic of
conditions in the accretion column.
\end{abstract}

 
\section{Introduction}
The accretion shock near the surface of a white dwarf in a magnetic
cataclysmic variable (MCV) will consist of a highly ionized \sqig10 \kev\
plasma cooling by bremsstrahlung emission. Most elements will be fully
ionized, leaving iron as the dominant cause of line emission. While model
fits to the X-ray spectra of MCVs have traditionally included an iron
\kalpha\ line (e.g.~Norton, Watson \&\ King 1991), it is only with the
\asca\ satellite that the data have sufficient spectral resolution to
investigate the structure of the line. 

An analysis of the \asca\ data on AO~Psc (Hellier \etal\ 1996) showed that
the thermal components of the iron \kalpha\ line were broadened by \sqig150
\ev.  However, reports on two other MCVs with the same detectors  [Kallman
\etal\ (1996) on BY~Cam and Fujimoto \&\ Ishida (1997) on EX~Hya] found a
\kalpha\ complex compatible with narrow emission lines. We have therefore
investigated the line widths in the class as a whole. In this paper we
present a systematic analysis of lines in 15 MCVs, and discuss the
mechanisms responsible for line broadening. Of our sample, AM~Her is a
phase-locked system (i.e.~a polar; see Cropper 1990 for a review), BY~Cam is
nearly phase-locked, and the remaining 13 are non-phase-locked systems
(i.e.~intermediate polars; reviewed by Patterson 1994).

\section{Data analysis}
The temperature of the highly-ionized plasma in an accretion shock will
typically range from 0.1--30  \kev. At the hottest temperatures iron will be
completely ionized or hydrogen-like (Fe {\sc xxvii} or {\sc xxvi}), and
give rise to a \kalpha\ line at 6.97 \kev\ (e.g.~Makishima 1986). At
temperatures of a few \kev\ iron will be helium-like ({\sc xxv}), producing
a \kalpha\ line at 6.70 \kev. Cold iron can produce a fluorescent line at
6.41 \kev. The gap between 6.41 and 6.70 \kev\ could be filled by
fluorescence from iron in intermediate states, but no iron species emits
between 6.70 and 6.97 \kev. 

We have used data from the \asca\ SIS0 and SIS1 instruments only (see 
Tanaka, Inoue \&\ Holt 1994), since the GIS instruments have insufficient 
resolution to separate the lines.  A list of observations analysed is given 
in Table~1. For all cases we used the same, fairly lax, screening
parameters, in order to maximise the available photons.  We analysed the
data in {\small BRIGHT2} mode where that was available for an entire
observation (XY~Ari, V1223~Sgr, TX~Col, EX~Hya, BY~Cam, RX\,1238--38) and in
{\small BRIGHT} mode otherwise. The resolution of the SIS instruments has
declined through the mission owing to radiation damage, so we computed
calibration matrices appropriate for each observation using 
{\small FTOOLS} v4.0.

\begin{table}\caption{{The \asca\ observations of MCVs. The time
is the amount of good data after screening.}}
\begin{tabular}{llc} \\
Star  &\ \ Date\ \ & Time (ks) \\ [3mm]
FO~Aqr & 1993 May 20 & 35 \\ [1mm]
EX~Hya & 1993 Jul 16 & 36 \\ [1mm]
AM~Her & 1993 Sep 27 & 39 \\ [1mm]
BY~Cam & 1994 Mar 11 & 48 \\ [1mm]
V1223~Sgr & 1994 Apr 24 & 59 \\ [1mm]
AO~Psc & 1994 Jun 22 & 81 \\ [1mm]
TX~Col & 1994 Oct 03 & 39 \\ [1mm]
PQ~Gem & 1994 Nov 04 & 77 \\ [1mm]
GK~Per & 1995 Feb 04 & 39 \\ [1mm]
TV~Col & 1995 Feb 28 & 38 \\ [1mm]
XY~Ari &  1995 Aug 06 & 34 \\ [1mm]
RX\,1712--24 & 1996 Mar 18 & 84 \\ [1mm]
BG~CMi & 1996 Apr 14 & 82 \\ [1mm]
V405~Aur {\small (0558+53)}& 1996 Oct 05 & 74 \\ [1mm]
RX\,1238--38 & 1997 Jan 14 & 58 
\end{tabular}\end{table}

We then analysed the spectra using {\small XSPEC} v9. However, since the
6.70 and 6.97 \kev\ lines are themselves blends of many components, we 
first needed to find the expected natural width, as seen by our
instrumentation. To do this we used {\small XSPEC}'s {\small CEMEKL} model
(Mewe, Kaastra \&\ Liedahl 1995) to predict the spectrum for a
multi-temperature plasma having emission measures distributed with a
power-law index of 1, up to a maximum temperature of 20 \kev. We then used
the  {\small FAKEIT} utility to create model spectra with the same photon
noise and spectral resolution as our datasets. We found that the resulting
6.70 and 6.97 \kev\ lines could be adequately fit with Gaussians. The
Gaussians had a best-fitting width of zero, and mean energies of 6.676 and
6.960 \kev\ respectively. For a fake dataset modelled on the AO~Psc data
from 1994 June the Gaussian width ($\sigma$) was less than 40 \ev\ (1
$\sigma$ error) or 60 \ev\ (90 per cent confidence). A fake dataset modelled
on the \rxs\ data of 1996 March gave larger upper limits of 60 and 85 \ev\
respectively, owing to the degradation in resolution. Since intermediate
polars often have highly absorbed spectra we have repeated the above
procedure with partial-covering absorbers (up to 5\pten{23}\ cm\mintwo )
included in the model. This made essentially no difference to the measured
line widths.

Turning to the observations, we fitted the summed spectra from SIS0 and SIS1 
simultaneously. To model the continuum we first fitted the energy ranges
4.0--6.0 and 7.5--10.0 \kev\ with a bremsstrahlung together with simple and
partial absorption. We then fixed the continuum parameters, extended the
fitted range to 4.0--10.0 \kev, and added Gaussians to model the iron lines.

We found that the iron lines of all observations could be modelled
satisfactorily by 3 Gaussians, corresponding to cold, helium-like and
hydrogen-like iron. Although one could argue for the inclusion of iron
emission between 6.41 and 6.70, and for an iron edge at 7 \kev\ in addition
to that introduced by the absorption, none of the observations
required their presence, so they were not included. Next, since many of the
observations had only barely enough S/N for our purpose, we constrained
the model by fixing the line energies to 6.41, 6.676 and 6.960 \kev\ (as
measured from the faked data). We also forced the two thermal lines to 
have the same width and fixed the width of the fluorescent
(6.41 \kev) line to zero, since none of the observations indicated that it
was broad. Thus we fitted only 3 line normalisations and one width.

\begin{table*}\caption{{Widths and equivalent widths of \kalpha\ iron lines 
in magnetic cataclysmic variables. The maximum, minimum and error values 
are 1-$\sigma$ limits. The last column gives the nominal resolution of
the SIS-1 detector at 6.5 \kev\  at the time of each observation (that for
the SIS-0 is marginally better).}}
\begin{tabular}{lccc@{\hspace{7mm}}ccccc} \\
Star & \multicolumn{3}{c}{{\hspace{-2mm}Equivalent widths (\ev)}} & 
\multicolumn{3}{c}{{\hspace{-5mm}Thermal line width ($\sigma$ in \ev)}} 
& \chisq & SIS-1 res \\ [1mm]
 & 6.41 \kev & 6.70 \kev & 6.97 \kev &\ \ Min\ \ &\ \ Best\ \ &\ \ Max\ \ 
& ($\nu \sim 300$) & ($\sigma$ in \ev) \\ [2mm]
V405~Aur & $65\pm 25$ & $380\pm 60$ & $0^{+40}$  &
300 & 450 & 530 & 0.95 & 110 \\ [1mm]
PQ~Gem & $100\pm 30$ & $50^{+90}_{-50}$ & $120^{+150}_{-120}$ &
190 & 330 & 620 & 0.98 & 87 \\ [1mm]
AO~Psc & $100\pm 40$ & $220\pm 90$ & $100\pm40$  &
180 & 250 & 280 & 1.06 & 81 \\ [1mm]
BG~CMi & $95\pm 30$ & $0^{+40}$ & $300\pm70$  &
175 & 230 & 332 & 1.02 & 105 \\ [1mm]
RX\,1712--24 & $65\pm 20$ & $160\pm 40$ & $85\pm30$  &
170 & 220 & 260 & 0.91 & 104 \\ [1mm]
TV~Col & $70\pm 30$ & $170\pm 70$ & $175\pm50$  &
150 & 200 & 250 & 0.96 & 91 \\ [1mm]
RX\,1238--38 & 0$^{+70}$ & $260\pm 60$ & $120\pm 75 $ &
100 & 160 & 210 & 1.06 & 114 \\ [1mm]
EX~Hya & $10^{+12}_{-7}$ & $390\pm 25$ & $110\pm15$  &
54 & 62 & 80 & 1.00 & 65 \\ [1mm]
AM~Her & $145\pm 15$ & $175\pm 20$ & $180\pm20$  &
53 & 67 & 88 & 1.07 & 69 \\ [1mm]
FO~Aqr & $140\pm 20$ & $90\pm 20$ & $85\pm20$  &
30 & 55 & 85 & 1.07 & 62 \\ [1mm]
XY~Ari & $10^{+25}_{-10}$ & $230\pm 45$ & $0^{+15}$  &
0 & 44 & 95 & 0.95 & 97 \\ [1mm]
V1223~Sgr & $105\pm 10$ & $75\pm 10$ & $80\pm10$  &
0 & 34 & 58 & 1.04 & 78 \\ [1mm]
TX~Col & $100\pm 40$ & $150\pm 50$ & $70\pm55$  &
0 & 0 & 100 & 0.97 & 85 \\ [1mm]
GK~Per & $50\pm 20$ & $70\pm 25$ & $0^{+20}$  &
0 & 0 & 86 & 0.90 & 90 \\ [1mm]
BY~Cam & $100\pm 20$ & $70\pm 20$ & $130\pm35$  &
0 & 0 & 69 & 0.99 & 76 
\end{tabular}\end{table*}

The resulting line widths and equivalent widths are given in Table~2. 
We find that 8 of the stars have thermal lines consistent
with zero width, but the remaining 7 stars have broad thermal lines
at $>90$ per cent confidence. We should caution that this analysis is
near the limit of the current data. For instance the result for BG~CMi,
no detection at 6.70 \kev\ but strong emission at 6.97 \kev, is physically
unlikely and implies an insecure result. Note also that while the results
for AO~Psc are qualitatively similar to those from the same data presented
in Hellier \etal\ (1996), in that both require broad thermal lines, the
fitted parameters are formally inconsistent. This can be explained by
different screening parameters, updated calibration files, and different
constraints on the iron line model, but still indicates the uncertainties 
in the analysis. A further effect is the progressive radiation damage
to the SIS detectors, complicating the comparison of different observations.
However, this is included in the response matrices, and as can be seen 
from the values in Table~2, the detected line broadening is in many cases
large compared to the resolution. 

Thus, while an individual result is uncertain, the finding  of broad lines
in roughly half the systems studied suggests that they  are a reality in
MCVs. To illustrate the findings, Fig.~1 shows the data and fitted model for
two of the systems with narrow lines and two of those with broad lines. 

We have also attempted to find limits on the width of the 6.41 \kev\
fluorescent line (which was fixed at zero in the above analysis). As a free
parameter we found it to be consistent with zero width in all cases. The
interesting case would be if we could prove the fluorescent line to be
narrower than the thermal lines. Because of the limited data quality this
was possible only in the case of AO~Psc, where the upper limit on the
fluorescent line width was 98 \ev\ (1 $\sigma$) or 129 \ev\ (90 per cent
confidence), compared to a 1 $\sigma$ lower limit of 180 \ev\ for the
thermal lines (Table~2). This difference is consistent with a separate 
origin for the fluorescent line (reflection from the white dwarf) 
compared to the thermal lines (see below).

\begin{figure*}\vspace{15cm}   
\caption{The SIS spectra of 4 MCVs, the upper two showing narrow
emission at 6.41, 6.70 \&\ 6.97 \kev, and the lower two showing 
broad thermal lines. Fitted parameters are in Table~2.}
\includegraphics{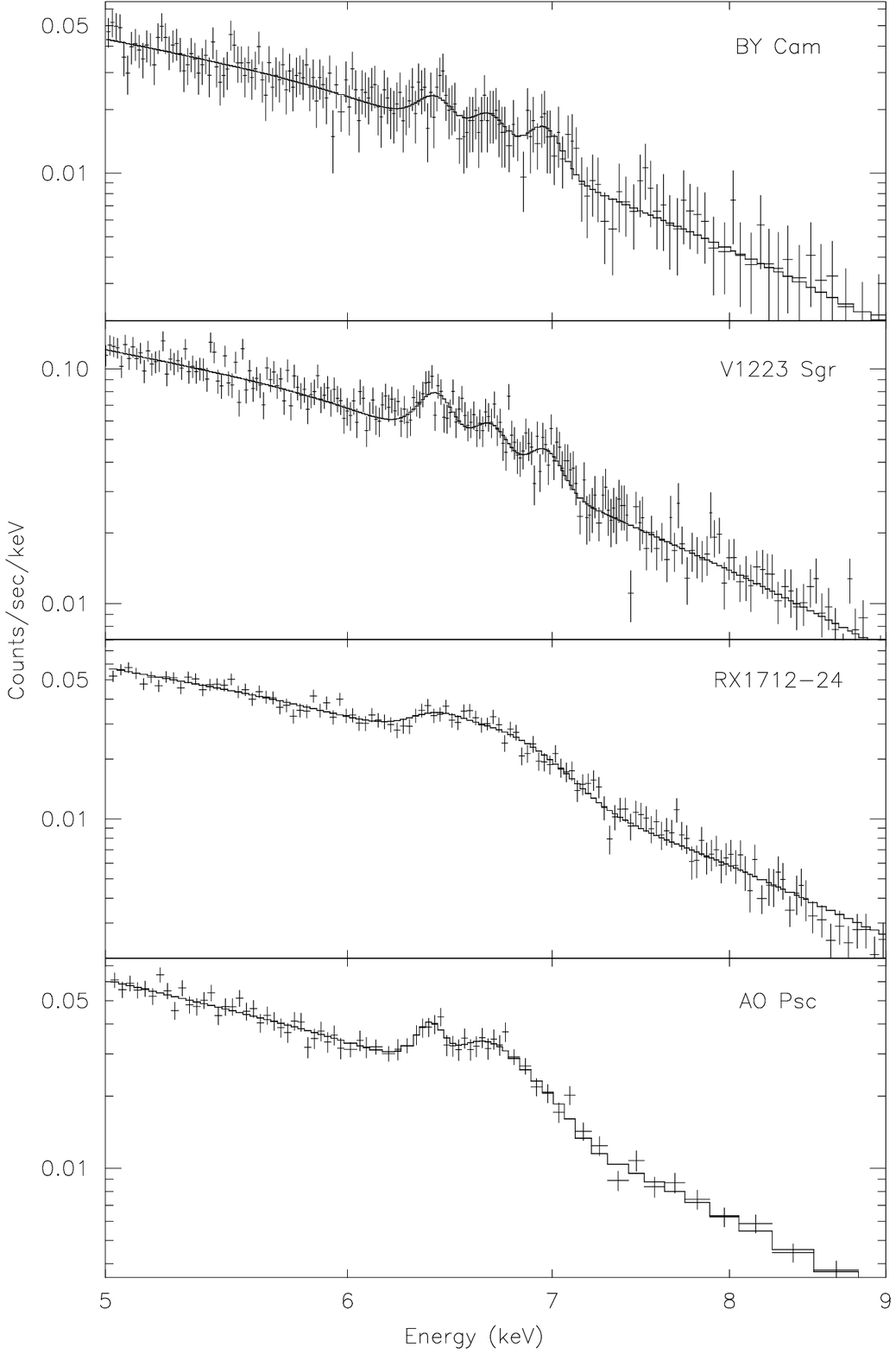}
\end{figure*}

\section{Broadening mechanisms}
Of the possible line broadening mechanisms, only two are likely to be
significant at the level of $\Delta E/E \sim 0.04$ seen in our data. These
are Doppler broadening and Compton scattering.

\subsection{Doppler Broadening}
Doppler broadening through the thermal motion of the ions in the plasma will
be of order $\Delta E/E \sim (2kT/m c^{2})^{1/2}$, which for our values is
typically 4 \ev\ and so not significant. 
However, the infalling material will hit the accretion shock at a speed near
the white dwarf escape velocity, which is 4000--16\,000 \kmps\ for white 
dwarfs of 0.5--1.4 \msun. The speed decreases by a factor of 4 in the shock
(e.g.~Frank, King \&\ Raine 1992) so that the X-ray emitting plasma will
have a bulk motion of 1000--4000 \kmps. This would produce a Doppler shift
of 20--90 \ev. The projection of this onto the line of sight would depend on
the geometry and would also vary with the white dwarf spin phase, so the
observed values would be lower still. Further, we would not see emission
with large blueshifts, since approaching material which was close to the
surface would be hidden by the white dwarf. We might see blueshifted photons
reflected towards us by the white dwarf, but these would be Compton
down-scattered by an amount  comparable to their blueshift (see next
section). Thus, we expect Doppler broadening to have a significant effect,
but overall it appears insufficient to explain the results of Table~2. 

\subsection{Compton scattering}
Roughly half of the X-rays from the accretion column will be directed towards
the white dwarf surface, and some will be reflected back towards us. Done,
Osborne \&\ Beardmore (1995) and Beardmore \etal\ (1995) find evidence for a
reflection component in the X-ray continua of MCVs, and propose reflection
from the white dwarf as the most likely cause of the 6.4 \kev\ fluorescent
emission seen in their data and in the tabulation of Table~2. 

The reflected \kalpha\ photons will be Compton down-scattered by up to 2
Compton wavelengths, depending on the scattering angle (e.g.~Rybicki \&\
Lightman 1979; Done \etal\ 1995). Thus the 6.70 and 6.97 \kev\ lines will
develop wings extending for 170 \ev\ to lower energies. These will have
equivalent widths of \sqig 10 per cent of the incident line (e.g.~Done
\etal\ 1995). Thus, although the broadening is of the right magnitude to
explain our results, and the 6.4 \kev\ emission shows that it must be
occurring, the inefficiency of reflection suggests that it would have a
relatively small effect on the overall line profile. Further, if the line
broadening were dominated by the reflection effect we might expect that
systems with broad lines had stronger 6.41 \kev\ emission, however
Table~2 shows no such effect. Note, though, that the equivalent widths of
the \kalpha\ lines are poorly understood, with smaller than expected fluxes
indicating either an underabundance of iron or optically thick lines (Done
\etal\ 1995).

We thus consider Compton scattering of the line photons within the accretion
column. For photons from a point source scattered once by electrons in a 
surrounding \sqig 5 \kev\ cloud the profile is broadened to $\Delta E/E \sim
(kT/m_{e} c^{2})^{1/2}$ or \sqig 0.1 (e.g.~Sunyaev 1980; Pozdnyakov, Sobol
\&\ Sunyaev 1983). Thus the broad lines could be explained if the column has
an optical depth sufficient to scatter most of the photons once, but not so
great that the lines are destroyed by multiple scattering. Resonant line
trapping enables this, since the cross-section to resonant scattering of
\kalpha\ photons is \sqig450 times that of Thomson scattering at 5 \kev\
(e.g.~Pozdnyakov \etal\ 1983; Matt, Brandt \&\ Fabian 1996). Thus if the
continuum is optically thin, the K$\alpha$ photons in the line core can be
resonantly trapped, but will escape immediately once they are Compton
scattered out of the core. For continuum optical depths of 0.05--0.2 \sqiglt
$\tau$ \sqiglt 1 almost all photons will be Compton scattered once and only
once (Illarionov \etal\ 1979; Pozdnyakov \etal\ 1983).

For comparison we expect an accretion column to have a dense, optically-thick
base at a temperature $<$\,1\kev. It decreases in density and optical 
depth while increasing in temperature as one moves up the column to an 
optically thin \sqig 20 \kev\ accretion shock (e.g.~Aizu 1973; Frank \etal\
1992). We can therefore envisage several regimes in the column: the
optically thick base produces no line emission; at continuum optical depths
of \sqig0.1--1 the line photons emerge with a singly-scattered profile
characteristic of the local temperature; higher up the column the
Compton-scattered component drops to insignificance once the column becomes
optically thin in the line core, and from here upwards the line emission is
narrow.

The temperatures at which these transitions occur are poorly known, but
could account for the difference in line widths in different systems. If the
$\tau = 0.1 - 1$ regime occurs where the plasma is sufficiently hot to emit
\kalpha\ photons (\sqiggt 3 \kev) we would expect  the broadened lines from
this region to dominate the \kalpha\ profile. Superimposed on it would be
weaker, narrow emission from the less dense, optically thin region, which
would reduce the overall width of the observed  profile. If, in contrast,
the transition from optical thickness occurs at a lower temperature
(\sqiglt 3 \kev), so that regions hot enough to  emit \kalpha\ are all
optically thin in the continuum, we would expect to see narrow lines.

In principle one could use the measured line widths to deduce the
temperatures of these transitions in each object. However, apart from
the limitations of the data, this requires simulations of Compton-scattered
profiles summing over the range of temperatures, densities and optical
depths in an accretion column model (e.g.~Aizu 1973), rather than
the current single-temperature simulations (e.g.~Pozdnyakov \etal\ 1983).
Further, published simulations concentrate the source photons into a point,
rather than distributing them through the column, a difference shown to be 
substantial by Sunyaev \&\ Titarchuk (1980). The addition of optically
thinner regions in the outer parts and higher up the column would help to
reduce the theoretical broadening of $\Delta E/E \sim 0.1$ at 5 \kev\
(e.g.~Sunyaev 1980) to the observed $\Delta E/E \sim 0.04$ in our data.

A last complication is the shape of the accretion column, expected to
have an arc-shaped footprint and to extend vertically, and its varying
projection onto the line of sight. The line photons will escape
preferentially in the direction of least scattering depth (e.g.~Swank,
Fabian \&\ Ross 1984) causing line flux changes as the white dwarf spins.
Further, changes in the line-of-sight optical depth round the spin
cycle could cause the temperature which corresponds to an optical depth
of \sqig 1 to vary, and so produce phase-dependent changes in the observed
line widths. These would be undetectable in current data, through lack
of photons, but are worth looking for with future missions.

\section{Discussion and Conclusions}
From the above discussion we conclude that the broad \kalpha\ iron lines
found in roughly half of the MCVs studied are caused by a mixture of
Doppler broadening due to radial infall, Compton down-shifted line emission
from a reflected component, and, probably most importantly, Compton
scattering of the line emission in the accretion column.  
A significant optical depth to scattering in the column may also be required to 
explain the line fluxes (e.g.~Swank \etal\ 1984; Done \etal\ 1995).

We have suggested that the broadening originates in the transition region
between optical thickness and optical thinness; that systems in which this
transition occurs at a temperature too low for significant \kalpha\ 
emission have narrow lines; and that systems with broad lines must have
regions of column which are still optically thick at a higher temperature 
(\sqiggt 3 \kev). 

We can test this by comparison with other work, taking the well
studied stars EX~Hya, as an example of a system with narrow lines,
and AO~Psc, the system with the clearest line broadening.
In EX~Hya the presence of lines of elements with
a lower ionization than iron implies a transition to optical thickness
at $<$\,1 \kev\ (Fujimoto \&\ Ishida 1997). In AO~Psc
the line ratios imply a higher temperature transition [Fujimoto \&\
Ishida (1995) quote $<$ 3 \kev, although this estimate is less certain
due to the weaker lines] in agreement with the above reasoning. 

Further, we can detect optical thickness in the accretion column by 
looking at changes in the X-ray continuum as the white dwarf spins. From
spin-resolved \asca\ spectroscopy of AO~Psc, Hellier \etal\ (1996) found
that the column contained several phases of absorption. The densest,
affecting regions of the column emitting at energies of at least 8 \kev,
requires an electron scattering column which changes by 6\pten{23}\
cm\mintwo\ over the spin cycle, suggesting an actual column of \sqig2\pten{24}\
cm\mintwo, and thus an optical depth of \sqig 1. This column is compatible
with an accretion rate of 10\up{17}\ g s\minone\ and an accretion area
covering $10^{-3}$ of the white dwarf surface (Hellier \etal\ 1996), values
which are in line with current estimates for intermediate polars
(e.g.~Patterson 1994; Hellier 1997). In contrast, EX~Hya shows much less
absorption, and in \asca\ data has no spin modulation above 6 \kev\ (Ishida,
Mukai \&\ Osborne 1994; Allan, Hellier \&\ Beardmore 1998), implying that it
is optically thin throughout the hard X-ray emitting regions. 

We can extend this difference into a test of our explanation, and predict
that the other systems with clearly narrow lines, such as V1223~Sgr, will
also be optically thin in the hard X-ray emitting regions. Thus 
spin-resolved spectroscopy with \asca\ would not show
a counterpart of the very dense absorber revealed in AO~Psc.
Unfortunately the test isn't clear cut
since we need to distinguish between a flux reduction caused by an optically
thick accretion column, and a flux reduction caused by emitting regions
passing over the limb of the white dwarf. However, at any one pole,
occultation effects and absorption effects are likely to occur in anti-phase
in MCVs (e.g.~Hellier, Cropper \&\ Mason 1991) allowing the test to be
performed given \asca -quality spectroscopy and an understanding of the spin
pulse in each system. Analysis of X-ray spectroscopy of other MCVs is thus
needed to confirm these ideas.

\section*{Acknowledgments}
We thank Andrew Beardmore and Chris Done for 
comments on the manuscript.

\end{document}